\documentclass[twocolumn]{aastex631}

\usepackage[normalem]{ulem}
\usepackage{xcolor}

\def\lesssim{\mathrel{\hbox{\rlap{\hbox{\lower4pt\hbox{$\sim$}}}\hbox{$<$}}}}
\def\gtrsim{\mathrel{\hbox{\rlap{\hbox{\lower4pt\hbox{$\sim$}}}\hbox{$>$}}}}

\newcommand{\rps}{\ensuremath{r_{\rm PS}}}
\newcommand{\ips}{\ensuremath{i_{\rm PS}}}
\newcommand{\zps}{\ensuremath{z_{\rm PS}}}
\newcommand{\yps}{\ensuremath{y_{\rm PS}}}

\newcommand{\grizy}{\ensuremath{grizy_{\rm P1}}}

\newcommand{\izy}{\ensuremath{izy_{\rm PS}}}

\newcommand{\degree}{\mbox{$^\circ$}}

\newcommand{\msol}{\mbox{M$_{\odot}$}}

\newcommand{\kms}{\mbox{$\rm{km}\,s^{-1}$}}

\newcommand{\CaII}{Ca~{\sc ii}}
\newcommand{\CaI}{Ca~{\sc i}}
\newcommand{\NaI}{Na~{\sc i}\,D}

\def\UTFermi{09 October 2022 at 13:16:59.99 UT} 
\def\MJDFermi{59861.55347}   
\def\Hostz{0.151}     
\def\HostDist{718\,Mpc}     
\def\HostDistMod{39.28}     
  
\def\Av{4.223}
\def\Arps{3.497}                  
\def\Aips{2.590}                  
\def\Azps{2.036}                  
\def\Ayps{1.674}                  

\def\Artot{4.64}                 
\def\Aitot{3.44}                  
\def\Aztot{2.70}                  
\def\Aytot{2.22}                  

\shorttitle{Optical light curve of GRB~221009A}
\shortauthors{Fulton et al.}

\begin{document}

\title{The optical light curve of GRB~221009A: the afterglow and the emerging supernova}

\author[0000-0003-1916-0664]{M. D. Fulton}
\affil{Astrophysics Research Centre, School of Mathematics and Physics, Queen's University Belfast, BT7 1NN, UK}
\author[0000-0002-8229-1731]{S. J. Smartt}
\affil{Department of Physics, University of Oxford, Denys Wilkinson Building, Keble Road, Oxford OX1 3RH, UK}
\affil{Astrophysics Research Centre, School of Mathematics and Physics, Queen's University Belfast, BT7 1NN, UK}
\author[0000-0003-2705-4941]{L. Rhodes} 
\affil{Department of Physics, University of Oxford, Denys Wilkinson Building, Keble Road, Oxford OX1 3RH, UK}
\author[0000-0003-1059-9603]{M. E. Huber} 
\affil{Institute for Astronomy, University of Hawai'i, 2680 Woodlawn Drive, Honolulu, HI 96822, USA}
\author[0000-0002-5814-4061]{V. A. Villar} 
\affil{Department of Astronomy and Astrophysics, Pennsylvania State University, 525 Davey Laboratory, University Park, PA 16802, USA}
\affil{Institute for Computational \& Data Sciences, The Pennsylvania State University, University Park, PA, 16802, USA}
\affil{Institute for Gravitation and the Cosmos, The Pennsylvania State University, University Park, PA 16802, USA}
\author[0000-0001-8385-3727]{T. Moore}
\affil{Astrophysics Research Centre, School of Mathematics and Physics, Queen's University Belfast, BT7 1NN, UK}
\author[0000-0003-4524-6883]{S. Srivastav}
\affil{Astrophysics Research Centre, School of Mathematics and Physics, Queen's University Belfast, BT7 1NN, UK}
\author[0000-0003-4717-9119]{A. S. B. Schultz} 
\affil{Institute for Astronomy, University of Hawai'i, 2680 Woodlawn Drive, Honolulu, HI 96822, USA}
\author[0000-0001-6965-7789]{K. C. Chambers} 
\affil{Institute for Astronomy, University of Hawai'i, 2680 Woodlawn Drive, Honolulu, HI 96822, USA}
\author[0000-0001-9695-8472]{L. Izzo} 
\affil{DARK, Niels Bohr Institute, University of Copenhagen, Jagtvej 128, 2200 Copenhagen, Denmark}
\author[0000-0002-4571-2306]{J. Hjorth} 
\affil{DARK, Niels Bohr Institute, University of Copenhagen, Jagtvej 128, 2200 Copenhagen, Denmark}
\author[0000-0002-1066-6098]{T.-W. Chen} 
\affil{Technische Universit{\"a}t M{\"u}nchen, TUM School of Natural Sciences, Physik-Department, James-Franck-Stra{\ss}e 1, 85748 Garching, Germany}
\affil{Max-Planck-Institut f{\"u}r Astrophysik, Karl-Schwarzschild Stra{\ss}e 1, 85748 Garching, Germany}
\author[0000-0002-2555-3192]{M. Nicholl} 
\affil{Birmingham Institute for Gravitational Wave Astronomy and School of Physics and Astronomy, University of Birmingham, Birmingham B15 2TT, UK}
\author[0000-0002-2445-5275]{R. J. Foley} 
\affil{Department of Astronomy and Astrophysics, University of California Santa Cruz, 1156 High Street, Santa Cruz, CA 95060, USA}
\author[0000-0002-4410-5387]{A. Rest} 
\affil{Space Telescope Science Institute, 3700 San Martin Drive, Baltimore, MD 21218, USA}
\affil{Department of Physics and Astronomy, The Johns Hopkins University, Baltimore, MD 21218, USA}
\author[0000-0001-9535-3199]{K. W. Smith}
\affil{Astrophysics Research Centre, School of Mathematics and Physics, Queen's University Belfast, BT7 1NN, UK}
\author[0000-0002-1229-2499]{D. R. Young}
\affil{Astrophysics Research Centre, School of Mathematics and Physics, Queen's University Belfast, BT7 1NN, UK}
\author[0000-0002-9774-1192]{S. A. Sim}
\affil{Astrophysics Research Centre, School of Mathematics and Physics, Queen's University Belfast, BT7 1NN, UK}
\author{J. Bright} 
\affil{Department of Physics, University of Oxford, Denys Wilkinson Building, Keble Road, Oxford OX1 3RH, UK}
\author[0000-0002-0632-8897]{Y. Zenati}
\affil{Department of Physics and Astronomy, The Johns Hopkins University, Baltimore, MD 21218, USA}
\author{T. de Boer} 
\affil{Institute for Astronomy, University of Hawai'i, 2680 Woodlawn Drive, Honolulu, HI 96822, USA}
\author{J. Bulger} 
\affil{Institute for Astronomy, University of Hawai'i, 2680 Woodlawn Drive, Honolulu, HI 96822, USA}
\author{J. Fairlamb} 
\affil{Institute for Astronomy, University of Hawai'i, 2680 Woodlawn Drive, Honolulu, HI 96822, USA}
\author[0000-0003-1015-5367]{H. Gao} 
\affil{Institute for Astronomy, University of Hawai'i, 2680 Woodlawn Drive, Honolulu, HI 96822, USA}
\author[0000-0002-7272-5129]{C.-C. Lin} 
\affil{Institute for Astronomy, University of Hawai'i, 2680 Woodlawn Drive, Honolulu, HI 96822, USA}
\author{T. Lowe} 
\affil{Institute for Astronomy, University of Hawai'i, 2680 Woodlawn Drive, Honolulu, HI 96822, USA}
\author[0000-0002-7965-2815]{E. A. Magnier} 
\affil{Institute for Astronomy, University of Hawai'i, 2680 Woodlawn Drive, Honolulu, HI 96822, USA}
\author[0000-0001-8605-5608]{I. A. Smith} 
\affil{Institute for Astronomy, University of Hawai'i, 34 Ohia Ku St., Pukalani, HI 96768-8288, USA}
\author[0000-0002-1341-0952]{R. Wainscoat} 
\affil{Institute for Astronomy, University of Hawai'i, 2680 Woodlawn Drive, Honolulu, HI 96822, USA}
\author[0000-0003-4263-2228]{D. A. Coulter} 
\affil{Department of Astronomy and Astrophysics, University of California Santa Cruz, 1156 High Street, Santa Cruz, CA 95060, USA}
\author{D. O. Jones} 
\affil{Gemini Observatory, NSF's NOIRLab, 670 N. A'ohoku Place, Hilo, Hawai'i, 96720, USA}
\author[0000-0002-5740-7747]{C. D. Kilpatrick} 
\affil{Center for Interdisciplinary Exploration and Research in Astrophysics (CIERA), Northwestern University, 1800 Sherman Ave, Evanston, IL 60201, USA}
\author[0000-0002-1052-6749]{P. McGill} 
\affil{Department of Astronomy and Astrophysics, University of California Santa Cruz, 1156 High Street, Santa Cruz, CA 95060, USA}
\author[0000-0003-2558-3102]{E. Ramirez-Ruiz} 
\affil{Department of Astronomy and Astrophysics, University of California Santa Cruz, 1156 High Street, Santa Cruz, CA 95060, USA}
\author[0000-0003-3004-9596]{K.-S. Lee}
\affil{Department of Physics and Astronomy, Purdue University, 525 Northwestern Avenue, West Lafayette, Indiana 47907, USA}
\author[0000-0001-6022-0484]{G. Narayan}  
\affil{Department of Astronomy, University of Illinois at Urbana-Champaign, 1002 W. Green St., IL 61801, USA}
\affil{Center for Astrophysical Surveys, National Center for Supercomputing Applications, Urbana, IL, 61801, USA}
\author[0000-0002-9176-7252]{V. Ramakrishnan}  
\affil{Department of Physics and Astronomy, Purdue University, 525 Northwestern Avenue, West Lafayette, Indiana 47907, USA}
\author[0000-0003-1724-2885]{R. Ridden-Harper}  
\affil{School of Physical and Chemical Sciences - Te Kura Matū, University of Canterbury, Private Bag 4800, Christchurch 8140, New Zealand}
\author{A. Singh}  
\affil{Departamento de Ciencias Fisicas, Facultad de Ciencias Exactas, Universidad Andres Bello, Fernandez Concha 700, Las Condes, Santiago, Chile}
\author[0000-0001-5233-6989]{Q. Wang} 
\affil{Department of Physics and Astronomy, The Johns Hopkins University, Baltimore, MD 21218, USA}
%
\author[0000-0002-5105-344X]{A. K. H. Kong} 
\affil{Institute of Astronomy, National Tsing Hua University, Hsinchu 30013, Taiwan}
\author[0000-0001-8771-7554]{C.-C. Ngeow} 
\affil{Graduate Institute of Astronomy, National Central University, 300 Jhongda Road, 32001 Jhongli, Taiwan}
\author{Y.-C. Pan} 
\affil{Graduate Institute of Astronomy, National Central University, 300 Jhongda Road, 32001 Jhongli, Taiwan}
\author[0000-0002-2898-6532]{S. Yang} 
\affil{Department of Astronomy, The Oskar Klein Center, Stockholm University, AlbaNova, 10691 Stockholm, Sweden}
\author[0000-0002-5680-4660]{K. W. Davis} 
\affil{Department of Astronomy and Astrophysics, University of California Santa Cruz, 1156 High Street, Santa Cruz, CA 95060, USA}
\author[0000-0001-6806-0673]{A. L. Piro} 
\affil{The Observatories of the Carnegie Institution for Science, 813 Santa Barbara St., Pasadena, CA 91101, USA}
\author[0000-0002-7559-315X]{C. Rojas-Bravo} 
\affil{Department of Astronomy and Astrophysics, University of California Santa Cruz, 1156 High Street, Santa Cruz, CA 95060, USA}
\author[0000-0002-1154-8317]{J. Sommer} 
\affil{Universitäts-Sternwarte München, Fakultät für Physik, Ludwig-Maximilians Universität München, Scheinerstr. 1, 81679 Munich, Germany}
\affil{Astrophysics Research Centre, School of Mathematics and Physics, Queen's University Belfast, BT7 1NN, UK}
\author[0000-0002-0840-6940]{S. K. Yadavalli} 
\affil{Department of Astronomy and Astrophysics, Pennsylvania State University, 525 Davey Laboratory, University Park, PA 16802, USA}
\affil{Institute for Computational \& Data Sciences, The Pennsylvania State University, University Park, PA, 16802, USA}
\affil{Institute for Gravitation and the Cosmos, The Pennsylvania State University, University Park, PA 16802, USA}



\begin{abstract}
We present extensive optical photometry of the afterglow of GRB~221009A. Our data cover $0.9 - 59.9$\,days from the time of \textit{Swift} and \textit{Fermi} GRB detections. Photometry in $rizy$-band filters was collected primarily with Pan-STARRS and supplemented by multiple 1- to 4-meter imaging facilities. We analyzed the Swift X-ray data of the afterglow and found a single decline rate power-law $f(t) \propto t^{-1.556\pm0.002}$ best describes the light curve. In addition to
the high foreground Milky Way dust extinction along this line of sight, 
the data favour additional extinction to consistently model the optical to X-ray flux with optically thin synchrotron emission.  We fit the X-ray-derived power-law to the optical light curve and find good agreement with the measured data up to $5-6$\,days. Thereafter we find a flux excess in the $riy$ bands which peaks in the observer frame at $\sim20$\,days. 
This excess shares similar light curve profiles to the type Ic broad-lined supernovae SN~2016jca and SN~2017iuk once corrected for the GRB redshift of $z=\Hostz$ and arbitrarily scaled. This may be representative of a supernova emerging from the declining afterglow. We measure rest-frame absolute peak AB magnitudes of $M_g=-19.8\pm0.6$ and $M_r=-19.4\pm0.3$ and $M_z=-20.1\pm0.3$. If this is an SN component, then 
Bayesian modelling of the excess flux would imply explosion parameters of 
$M_{\rm ej}=7.1^{+2.4}_{-1.7}$\,M$_{\odot}$, 
$M_{\rm Ni}=1.0^{+0.6}_{-0.4}$\,M$_{\odot}$,
and 
$v_{\rm ej}=33,900^{+5,900}_{-5,700}$\,\kms, 
for the ejecta mass, nickel mass and ejecta velocity respectively, inferring an explosion energy of 
$E_{\rm kin}\simeq 2.6-9.0\times10^{52}$\,ergs.
\end{abstract}



\section{Introduction} \label{sec:intro}

Long-duration gamma-ray bursts (lGRBs) are typically associated with the signature of a broad-lined type Ic supernova (SN) in their light curves and spectra, as the afterglow fades and an SN rises within 10-20 days. Since the discovery of SN~1998bw/GRB~980425 
and SN~2003dh/GRB~030329 \citep{1998Natur.395..670G,2003Natur.423..847H}
more than 40 probable GRB supernovae have been observed 
\citep[e.g.][]{2012grb..book..169H,2017AdAst2017E...5C}.  However, a supernova
signature is not always detected for nearby lGRBs \citep{2006Natur.444.1050D,2006Natur.444.1047F}, 
leading to speculation that massive stellar deaths are not the source of all lGRBs \citep{2022ApJS..259...67L, 2022Natur.612..223R}.

GRB~221009A was first detected and announced by the Neil Gehrels \textit{Swift} Observatory via the Gamma-ray Coordinates Network (GCN) circulars \citep{dichiara2022swift}.  A hard X-ray source and an optical 
counterpart at Swift-UVOT white light magnitude 16.63 were reported, locating it within the Galactic plane at  
a latitude of $b=4.322^{\circ}$. The lGRB nature was confirmed by \cite{kenneaSwift22}, while the Gamma-ray Burst Monitor (GBM), on-board the \textit{Fermi} spacecraft, reported a detection 1hr before the Swift trigger time \citep{veres2022grb}, noting it was the brightest GRB ever detected by Fermi-GBM instrument. Fermi's Large Area Telescope (LAT) further 
reported the detection of the GRB and recorded its highest-energy photon at 7.8 GeV \citep{BissaldiFermiLAT22}. The first ground-based detections of the afterglow in the optical by \cite{perley2022swift} and in the radio by \cite{BrightAMILA22} were followed by a multitude of 
GCN circulars reporting measurements across the electromagnetic spectrum. 
Interest in this extraordinary event further increased
when the LHAASO experiment reported the detection of more than 5,000 very high energy (VHE)
photons with energies up to 18\,TeV \citep{LHAASO22}. The Carpet-2 experiment reported a 
possible 251\,TeV photon detection \citep{2022ATel15669....1D} which would be remarkable if proven reliable. Spectra of the afterglow were obtained by \cite{2022GCN.32648....1D} and \cite{2022GCN.32686....1C}, both of which 
reported a red continuum with absorption features that correspond to \CaII, \CaI, \NaI\ at a redshift of $z=\Hostz$. A host was later confirmed, and a similar redshift of $z=\Hostz$ was determined by \cite{2022GCN.32765....1I} through the identification of host galaxy absorption and emission lines. 
At this extragalactic distance, such VHE photons should be absorbed 
through pair-production when they scatter off the extragalactic background light, raising
the possibility of axion-like particle production \citep{2022arXiv221102010C,2022arXiv221007172B}. 
The remarkably high fluence, luminosity and detection of VHE photons make GRB~221009A an object
of broad interest.

In this Letter, we present the detection of a supernova
signature in the fading afterglow of GRB~221009A, the 
brightest GRB known to date. We present an extensive photometric data set 
primarily from the Pan-STARRS2 Observatory, supported by 
 multiple other facilities. The search for 
a supernova signal is complicated due to 
the high, and uncertain, foreground extinction,
the bright afterglow and the uncertainty in 
host galaxy contribution. 
Throughout this Letter we assume a GRB detection time of $T_0=\MJDFermi$ (\UTFermi) from \textit{Fermi} \citep{veres2022grb}, a Hubble Constant of $H_0=70$\,\kms\,Mpc$^{-1}$ and the 
redshift of $z=\Hostz$ \citep{2022GCN.32648....1D,2022GCN.32686....1C,2022GCN.32765....1I}. This corresponds to a luminosity distance of $D_{\rm L}=$\HostDist\ and distance modulus 
$\mu=\HostDistMod$ assuming a flat Universe with $\Omega_{\rm M}=0.3$. 

\section{Observational data} \label{sec:data}

The Pan-STARRS (PS) system comprises two 1.8\,m telescope units located at the summit of Haleakala on the Hawaiian island of Maui \citep{2016arXiv161205560C}. The Pan-STARRS1 (PS1) telescope is fitted with a 1.4 Gigapixel camera (GPC1) with 0\farcs26 arcsec pixels providing a 3.0 degree diameter focal plane, corresponding to a field-of-view area of 7.06 sq deg. The Pan-STARRS2 (PS2) telescope is fitted with a similar but larger 1.5 Gigapixel camera (GPC2) resulting in a slightly wider field-of-view. Both telescopes are equipped with an SDSS-like filter system, denoted as \grizy.  The Pan-STARRS1 Science Consortium (PS1SC) 3$\pi$ Survey produced $grizy_{P1}$ images of the whole sky north of $\delta = -30\degree$ \citep{2016arXiv161205560C}. Multi-epoch observations spanning 2009-2014 have been stacked and a public data release provides access to the images and catalogues \citep{2020ApJS..251....7F}. These data provide reference images for immediate sky subtraction, which allows the discovery of transients and accurate photometry with host galaxy removal. 
Pan-STARRS data from both telescopes are processed in real-time as described in \cite{2020ApJS..251....3M}, \cite{2020ApJS..251....6M} and \cite{2020ApJS..251....4W} at the University of Hawaii and the transient sources 
are selected, filtered and classified by the Transient Science Server at Queen's
University Belfast \citep{2020PASP..132h5002S}. In normal survey and discovery mode,
these feed science programs such as the Young Supernova Experiment \citep{2021ApJ...908..143J} and the 
Pan-STARRS Search for kilonovae \citep{McBrien2020}.

The Pan-STARRS afterglow observations of GRB22100A presented here were all taken with PS2. After the discovery of the GRB 
and the optical afterglow, 
we triggered PS2 to observe in 
\rps, \ips, \zps\ and \yps, with a nightly sequence between 11th October and 4th December 2022 \citep{2022GCN.32758....1H}, depending on Moon phase and weather.
Photometry was carried out on the difference images with the Pan-STARRS1 3$\pi$ sky survey data \citep{2016arXiv161205560C}
used as references. The 3$\pi$ data are made of stacks of short exposures
and the total exposure times and depths at the position of the afterglow 
(which we measure at RA$=288.26459^{\circ}$ DEC$=+19.77341^{\circ}$) 
are listed in Table\,\ref{tab:3Pistacks}.  The photometric zeropoints on the 
PS2 target images were set with the Pan-STARRS1 $3\pi$ catalogue \citep{2020ApJS..251....7F}. We typically used 100-200\,sec exposures and stacked
images on any one night (from 1 to 12 images, depending on target magnitude and sky brightness).

\begin{table*}
\centering
\begin{tabular}{cccccccc}
\hline
Filter   &  Exposure time & \multicolumn{2}{c}{3$\sigma$ depth} & $A^{MW}_{\lambda}$ &$A^{Tot}_{\lambda}$ & $\lambda_{\rm obs}$ & $\lambda_{\rm rest}$ \\
               & (sec)          & AB Mag & Flux ($\mu$Jy) & mag &  mag & (nm) & (nm)  \\
               \hline    
\rps   &   636           & $>22.24$ & $<4.61$ & \Arps  & \Artot & 617 & 536 \\  
\ips   &   930-1020      & $>22.01$ & $<5.70$ & \Aips  & \Aitot  & 752 & 653 \\  
\zps   &   540-570-600   & $>21.14$ & $<12.71$ & \Azps  & \Aztot  & 866 & 752  \\  
\yps   &   620           & $>20.34$ & $<26.54$ & \Ayps  & \Aytot  & 962 & 836 \\  
\hline
\end{tabular}
\caption{Depths of the 3$\pi$ reference images used for PS2 image data template subtraction. The effective wavelengths of the Pan-STARRS
filters are from \cite{2012ApJ...750...99T} and the corresponding restframe wavelengths at $z=0.151$ are listed. The Galactic foreground extinction ($A^{MW}_{\lambda}$) in each filter is from \cite{2011ApJ...737..103S} and the total extinction required for the observed X-ray to optical flux ratio is $A^{Tot}_{\lambda}$. } 
\label{tab:3Pistacks}
\end{table*}

Two methods were used to measure the flux on the difference images. A difference image was created from each individual exposure, and a point-spread-function (PSF) was forced at the GRB~221009A afterglow position (measured from the early, bright epochs). We statistically combined the measured PSF flux from each difference image through a weighted average, using a small temporal bin size of 0.125 days (3 hours). For the later Pan-STARRS epochs
 ($t-T_0 > 34$\,days) we increased the bin size to four days in the $zy$-filters to enhance the signal-to-noise. Alternatively, an image stack on each night was created, and a difference image produced from the stack. Again, a PSF was forced at the GRB afterglow position and flux measurement used.  All fluxes and magnitudes quoted here are in microJanksys ($\mu$Jy) and AB mags. 
 The results from image stacking were used instead of the weighted average of fluxes only when the object fell on a masked chip within the camera CCD which prevented the typical pipeline processing of target images described in \cite{2016arXiv161205560C}. 
 Regardless of the method, the resulting flux measurements were calibrated carefully to the Pan-STARRS1 DR2 3$\pi$ reference catalogue \citep{2020ApJS..251....7F} using approximately 1000 field stars visible within the target frames.

While most of the data here are provided by PS2, we gathered other important photometric 
data with the 0.4-meter SLT R-C Telescope \citep{2022GCN.32667....1C} and 1-meter LOT Cassegrain Telescope at the Lulin Observatory, Taiwan; the Dark Energy Camera (DECam) on the 4-meter Telescope at the Cerro Tololo Inter-American Observatory, Chile; the 4.1-meter Southern Astrophysical Research (SOAR) Telescope, Chile; the 1-meter Swope Telescope, Chile; the IO:O on the 2.2-meter Liverpool Telescope (LT), La Palma; and MegaCam on the 3.6-meter Canada-France-Hawaii Telescope, Hawaii.

Eight epochs of DECam observations were conducted between 16th October 2022 (MJD 59868.01) and 31st October 2022
(MJD 59883.011) taking between $2-5\times100$\,sec exposures in the filters $r$ and $i$. 
The data were reduced and photometrically calibrated with the {\tt photpipe} package
\citep{2014ApJ...795...44R} using the images from  
16th October 2022 as templates. These were subtracted from all subsequent images and a
PSF was forced at the afterglow position on the difference images. Since the template contains 
transient flux and we were not able to get a final set of templates in which the afterglow had 
faded, we applied an offset to match the DECam $r$-filter flux measured on MJD 59880.01 to the 
SOAR epoch on MJD 59880.02, and the DECam $i$-filter flux measured on MJD 59875.01 to the 
PS2 epoch on MJD 59875.23. These offsets were subsequently applied to all the DECam
difference images in the respective filters. Data from Swope were subjected to difference imaging, using the Pan-STARRS1 3$\pi$ references as templates and forced photometry was implemented thereafter. 


We used the three epochs of Hubble Space Telescope (HST) data that is publicly available through the 
DDT program of \cite{levan2022grb}. The WFC3 passbands of the F625W, F775W and F098M filters are similar to that of the \rps,\, \ips\, and \yps\, filters respectively (see Section\,\ref{sec:sn}). PSF magnitudes were measured on the HST target images using the DOLPHOT package \citep{2016ascl.soft08013D}. 
Observations conducted by SLT, LOT, SOAR, LT and MegaCam between 10th October and 8th November were not subjected to any form of image subtraction, instead, a PSF was forced onto the target images using python packages: \textit{Astropy} and \textit{Photutils} \footnote{Python tool used to measure PSF photometry can be found on GitHub. https://github.com/mnicholl/photometry-sans-frustration}
and the resulting flux measurements were calibrated against Pan-STARRS1 $3\pi$ survey field stars. 

Since no difference imaging was applied to the SLT,
LOT, SOAR or LT one may be concerned by late-time host-galaxy flux contamination. This is
particularly concerning when the measured flux of the transient is comparable to, or fainter than, the limits we can put on the host galaxy from the Pan-STARRS1 3$\pi$ data (see Table\,\ref{tab:3Pistacks}).  However the 
photometry between different instruments (with and without difference imaging)
is consistent within the statistical uncertainties. A probable faint 
host galaxy is 
visible in the deepest F625W and F775W HST images, approximately 0\farcs5 offset to the North-East. However, this is too faint to contribute significantly to the 
$r$ and $i$ photometry and our \yps\ data are all image subtracted. 
Hence we make the assumption that there is no host galaxy flux  
contributing to the non-differenced images in the filters $r, i$ or $z$.
This can only be confirmed with deep observations in the next observing season. We list our measurements also in fluxes (microjanskys) within Appendix\,\ref{app:photometrytable}
so that a future correction can be applied should that be necessary. 

\begin{figure}
    \centering
    \includegraphics[width = \columnwidth]{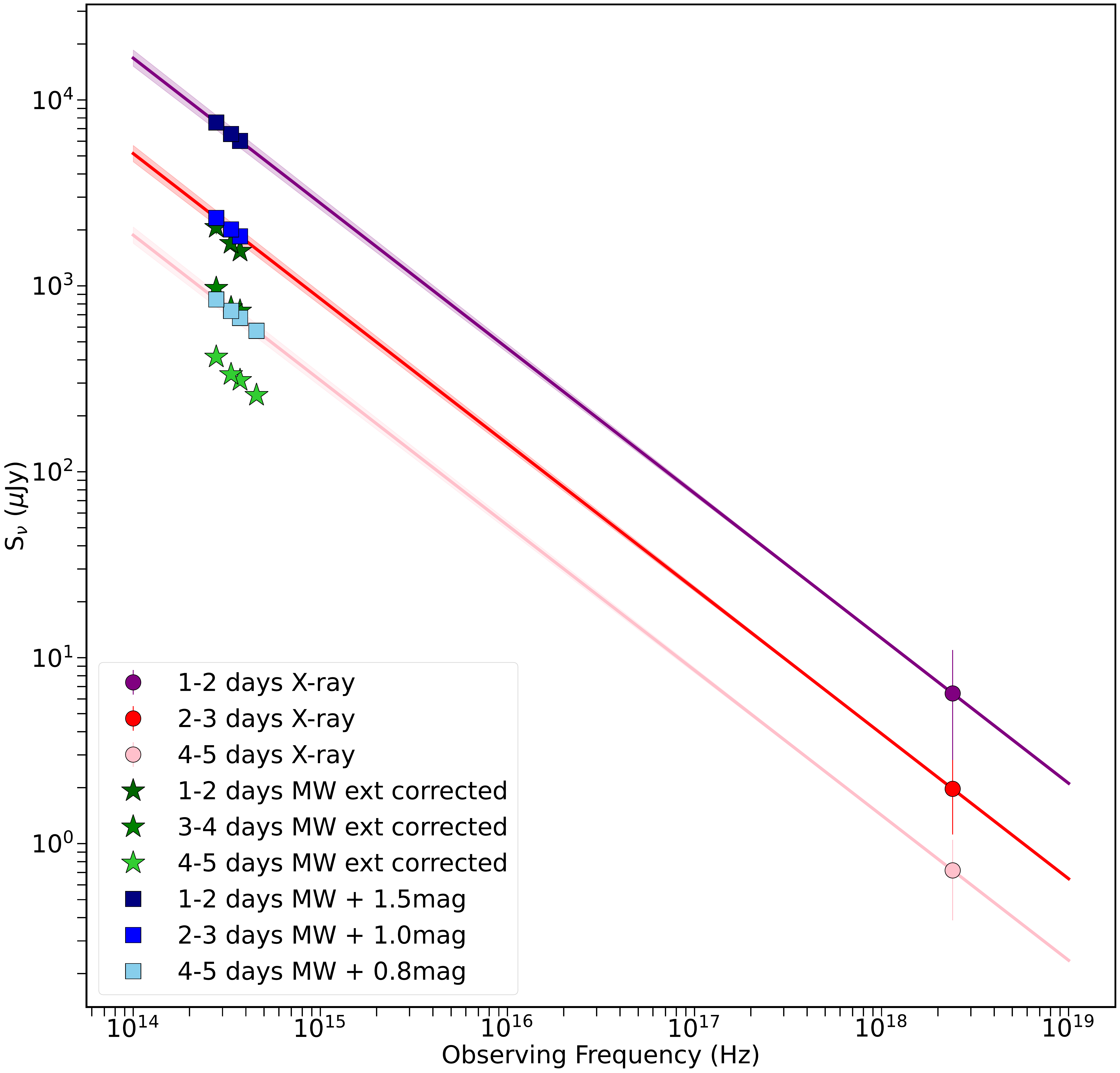}
    \includegraphics[width = \columnwidth]{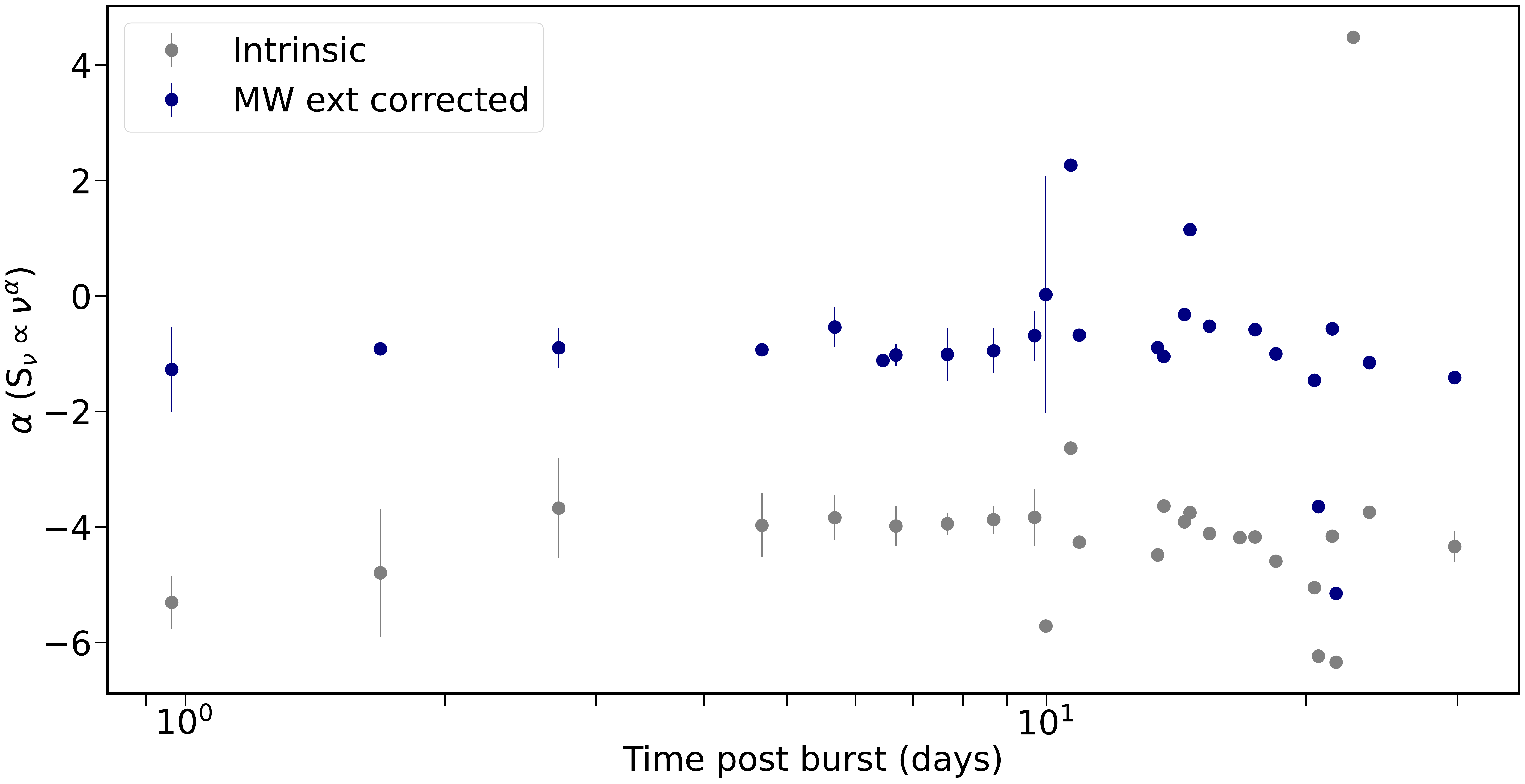}
    \caption{\textit{Upper Panel}: Comparison of the X-ray flux densities (and photon index) with the optical flux densities corrected for the known Milky Way extinction (stars) for three epochs: +1-2, +2-3 and +4-5 days post-burst. Also shown (squares) are the optical flux densities with an additional extinction correction such that it agrees with the X-ray spectrum. We estimate that at least another 0.8\,mag of extinction (averaged across the optical bands) is required along the line of sight in addition to what is provided by the Milky Way extinction maps \citep{2011ApJ...737..103S}.
    \textit{Lower panel}: spectral index fit to the optical bands at all wavelengths, both the observed and the Galactic extinction corrected points. Even after correcting for Galactic extinction, the optical spectral index is still too shallow to agree with the X-ray spectrum. This further suggests a need for an additional absorption component.}
    \label{fig:opt-xray}
\end{figure}

\section{Analysis of the x-ray and optical afterglow}
\label{sec:analysis}

The X-ray counterpart to GRB 221009A was observed by the Neil Gehrels \textit{Swift} Observatory X-ray telescope (XRT) starting 0.9\,hours after the \textit{Fermi} trigger \citep{veres2022grb}, and is still observing at the time of writing. We downloaded the \textit{Swift} XRT data to date from the \textit{Swift} Burst Analyser \citep{2007A&A...469..379E, 2009MNRAS.397.1177E, 2010A&A...519A.102E}. A single decaying power-law can best describe the XRT light curve. We fit a power-law component to the first 60 days of flux data, use the fluxes and not flux densities so as not to introduce any spectral bias, and found the light curve is described with $f(t) \propto t^{-1.556\pm0.002}$. There is no evidence of any breaks in the light curve which could result from either a spectral break (e.g. the cooling break) passing through the band or a jet break. There is also no evidence of a change in the X-ray photon index (related to the spectral index), indicating no significant spectral evolution occurring over the first 60 days post-burst. The \textit{Swift} Burst Analyser quotes an X-ray photon index of $1.78\pm0.01$ which corresponds to a spectral index ($S({\nu}) \propto \nu^{\alpha}$) of $\alpha = -0.78\pm0.01$. The measured X-ray light curve decay and spectral index indicate the X-ray emission originates from optically thin synchrotron radiation, where the synchrotron cooling break has a frequency that is higher than that of the observing band.

\begin{figure}
    \centering
    \includegraphics[width = 0.95\columnwidth]{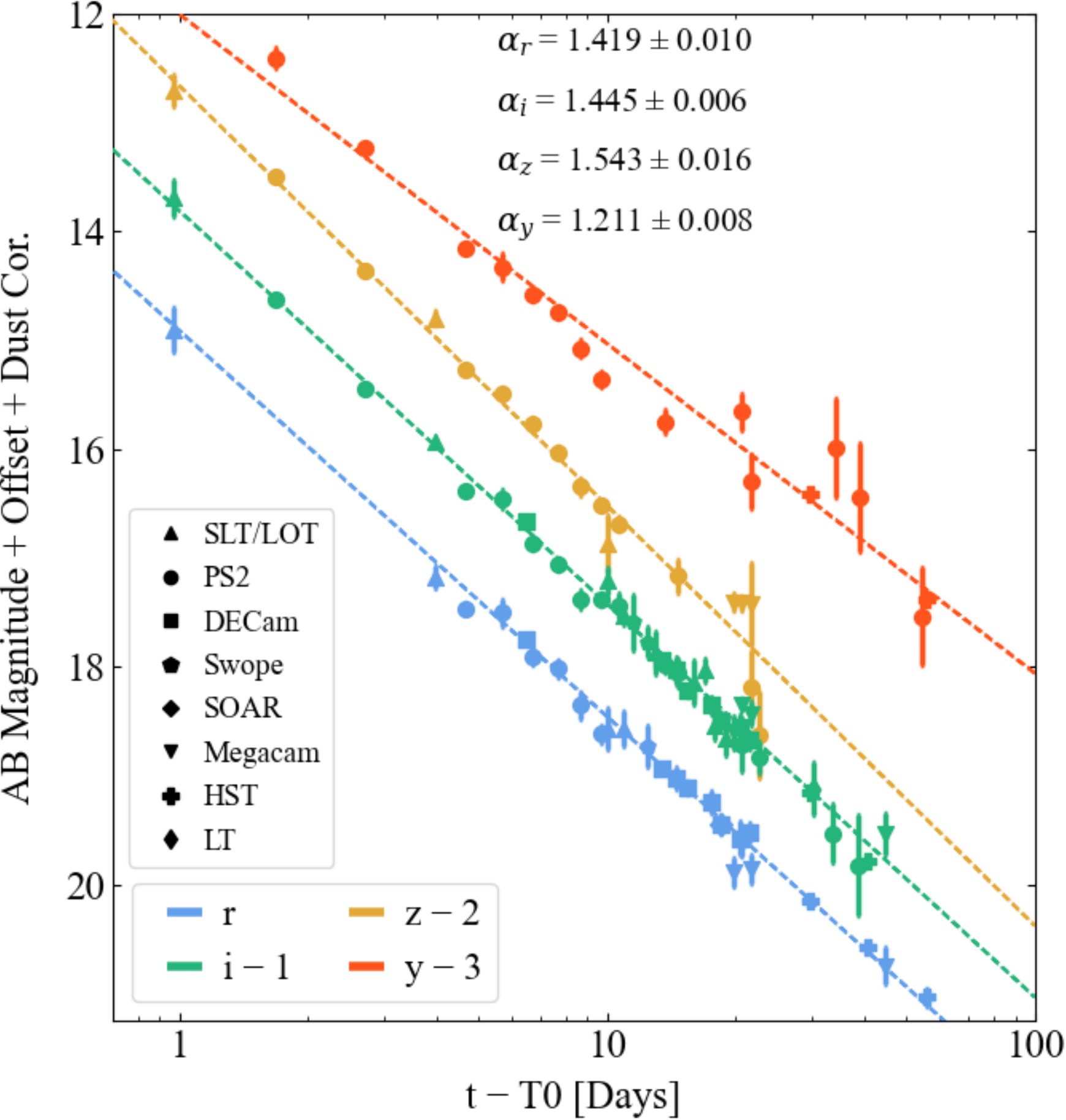}
    \includegraphics[width = 0.95\columnwidth]{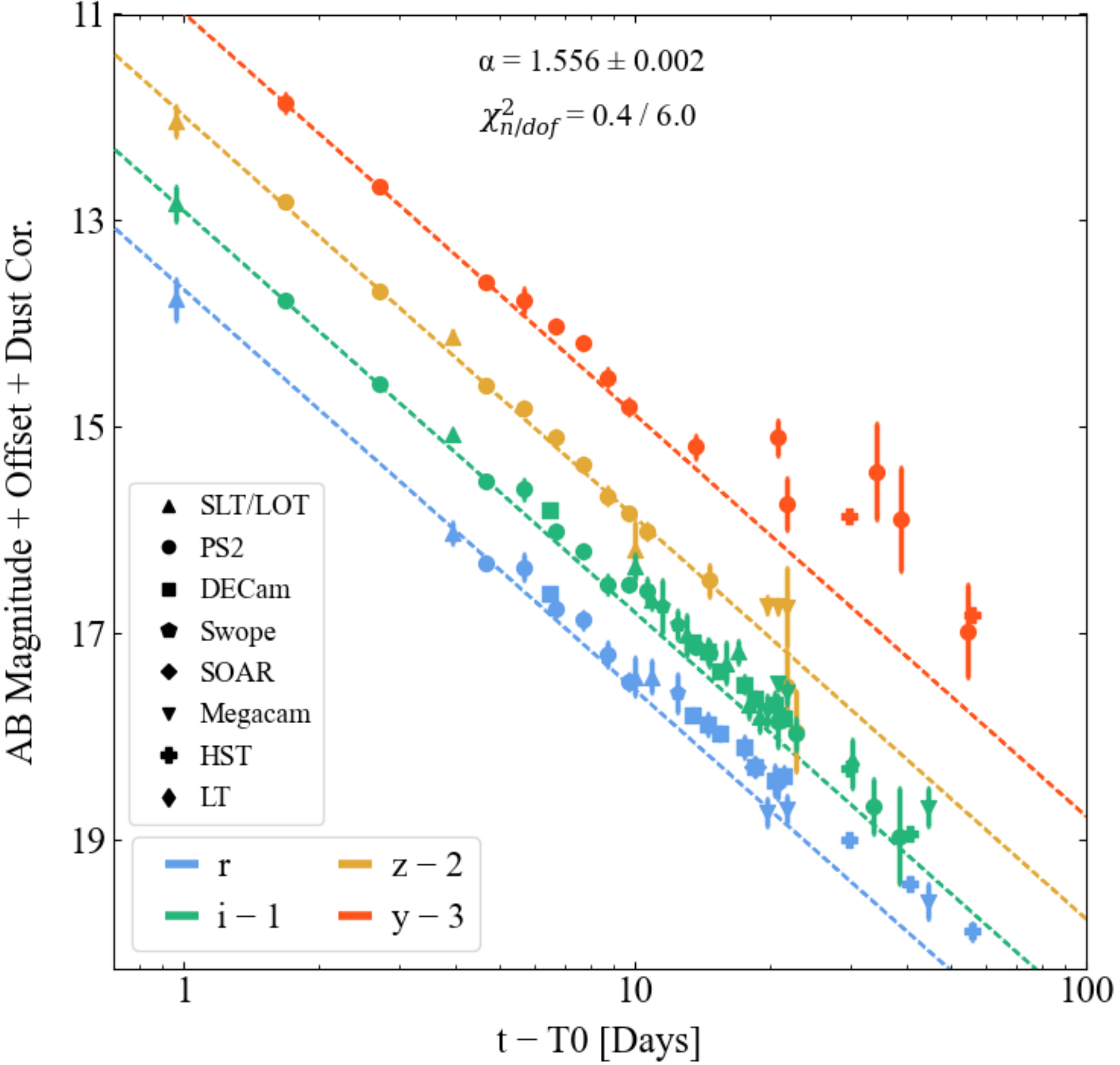}
    \caption{Upper Panel: the optical afterglow of GRB~221009A in the $rizy$-filters. A decaying power-law has been fit to each filter independently using the measurements across all epochs which have been corrected for galactic dust extinction. The data points are in the observer frame with the instruments and telescopes as in the legend. Only magnitude measurements with $\geq2\sigma$ significance are included. Bottom Panel: the same optical afterglow of GRB~221009A, this time with a single power-law of $f(t)\propto t^{-1.556}$ (derived from the X-ray measurements) fit to all filters using only data points up to seven days after GRB trigger which have been corrected for galactic dust extinction plus extra extinction required by X-ray analysis.}
    \label{fig:afterglow-lc}
\end{figure}

Given that the optical light curve is also showing a decay, and that only optically thin synchrotron radiation produces a decaying light curve within the fireball model, the optical emission should be on the same branch of the afterglow spectrum as the X-ray band and thus the decay rates should be identical \citep{1998ApJ...497L..17S, 2002ApJ...568..820G}. A power-law decay also best describes the measured optical decay. However, the difference between the optical and X-ray decay rates is not large enough to be consistent with the presence of a spectral break between the two bands.
The difference in light curve decay rates should be 1/4 if the cooling break is between the optical and X-ray bands and accompanied by a spectral index difference of 1/2 \citep{2002ApJ...568..820G}.

\begin{figure*}
    \centering
    \includegraphics[width = 0.90\textwidth]{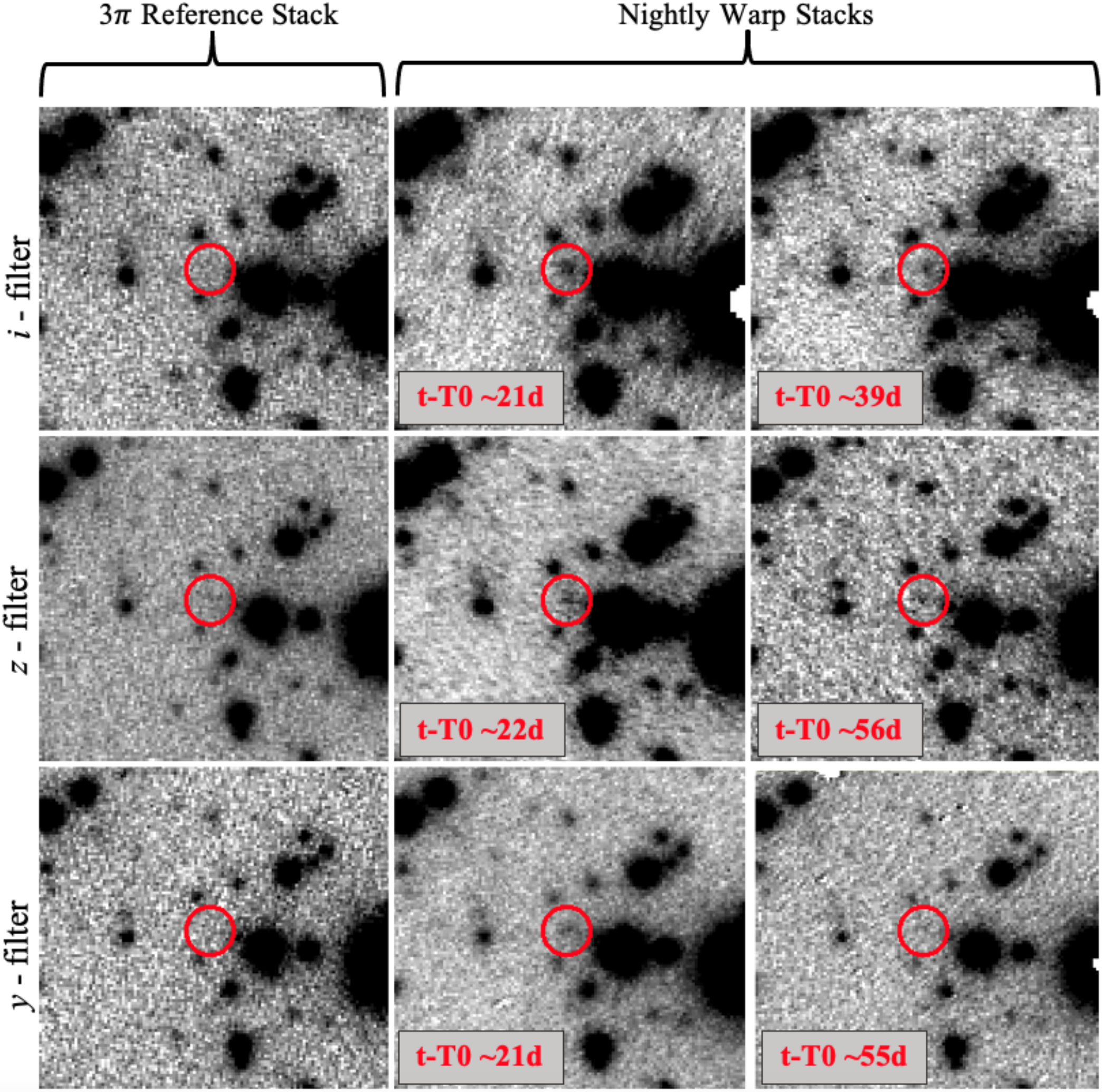}
    \caption{Pan-STARRS reference images and detections
        \textit{Top triplet}: \ips-filter reference image, intra-night stacks from MJD 59882 (+21d, around SN peak) and MJD 59900 (+39d). 
        \textit{Middle triplet}: \zps-filter reference image, intra-night stacks from MJD 59883 (+22d) and MJD 59917 (+56d).
        \textit{Bottom triplet}: \yps-filter reference image, intra-night stacks from MJD 59882 at (+21d, around SN peak), and MJD 59916 (+55d, object gone).
        Location of GRB~221009A is centered inside the red circle.
    }
    \label{fig:images}
\end{figure*}

In addition to differing light curve decay rates, the optical flux densities do not agree with what is predicted from the X-ray spectrum (Figure \ref{fig:opt-xray}). This is the result of 
the line of sight extinction through the Galactic plane being high at $A_V=\Av$
\citep{2011ApJ...737..103S}. We confirm that the extinction is high by measuring an optical spectral index of ($\alpha \approx -4$) which is notably steeper than the expected and measured X-ray spectral index of the synchrotron emission, 
$\alpha \approx -0.8$. We correct the optical data for the Galactic extinction across the specific $rizy_{\rm PS}$-filters as reported in the NASA Extragalactic Database \citep{https://doi.org/10.26132/ned5}, which uses the maps of 
\cite{2011ApJ...737..103S} and the \cite{1999PASP..111...63F} reddening law
(see Table\,\ref{tab:3Pistacks}). 
Figure \ref{fig:opt-xray} shows the X-ray to optical SED. Once corrected for Galactic extinction (only), the optical flux densities are still  too faint to agree with the extrapolation of the X-ray data using a spectral index of $S(\nu)\propto \nu^{-0.78\pm0.01}$. Additional 
extinction is required and the amount  required to reconcile the optical ad X-ray fluxes varies across the three epochs shown in Figure \ref{fig:opt-xray}.  The optical spectral index (Figure\,\ref{fig:opt-xray}) is also lower than the expected $\alpha\simeq-0.8$, further supporting a requirement for a source of extra extinction. 

The earliest epoch  (1-2\,days post-burst) requires 1.5\,magnitudes 
of extra extinction, reducing to 
1.0 (at 2-3\,days) and 0.8 magnitudes (4-5\,days). It's not clear why the implied extra extinction would 
vary across the three epochs but time-variable extinction has been proposed
before \citep[e.g. GRB~190114C;][]{2021A&A...649A.135C, 2022A&A...659A..39M}. 
We use the value  of 0.8 magnitudes 
(which is the average value across the $rizy_{\rm PS}$ bands) as this is the
closest value in time to the possible emergent SN signal. 
There are three possibilities. 
for this additional line of sight 
extinction. The Galactic dust structure may not be accurately captured by the low resolution Milky Way extinction maps of 
\cite{2011ApJ...737..103S} or it 
is 
possible that $R_V>3.1$, which may 
be plausible in this high density region. The third possibility is
 additional absorption in the host. 
 This may not be surprising given four of the five GRBs with very high energy detections ($E_\gamma>100$\,GeV) also have evidence of strong dust contamination as mentioned in \citet{2022MNRAS.513.1895R}. The VHE GRBs appear to have an increased likelihood of dust absorption compared to the rest of the long GRB population where only about 25\% of events have significant optical extinction \citep{2012ApJ...746..156C}. 

We plot our optical data in AB magnitudes in Figure\,\ref{fig:afterglow-lc}, including all points with $\geq$2-sigma significance. 
We first measured the optical decay rate in all filters independently and across all epochs ($T_0 + 1 < t \lesssim T_0 + 56$). The decay rate in the $ri$-filters follow $f(t) \propto t^{-1.43\pm0.02}$, the $z$-filter follows $f(t) \propto t^{-1.54\pm0.02}$ and the $y$-filter follows $f(t) \propto t^{-1.21\pm0.01}$. There isn't a single decaying term that can describe the fade in all filters, and, with the exception of the $z$-filter, the decay terms derived are substantially shallower than that of the X-ray slope.
Using epochs from ($1 \leq t-T_0 \leq 4.7$\,d) in the 
\izy\, filters, we also fit the X-ray derived slope of $f(t)\propto t^{-1.556\pm0.002}$ to these points. In the $\rps$-filter, we lack data at 2-3 days so we extend the temporal baseline to include the 6.7\,d PS2 point. We find the early data are well explained by the model, as evidenced by the measured normalized chi-squared value of $\chi^{2}_{n}/dof = 0.4/6.0$. 

The dissimilar decaying rates between the optical filters, and the deviation of the optical light curve from the X-ray light curve starting at $t-T_0 > 7$\,days are suggestive of another component appearing in the data. The light curve deviation is most pronounced in the \yps\ filter at $t-T_0 = 21$\,days, and we discuss this in the next section. 
Figure\,\ref{fig:images} illustrates the data quality and 
detections around the peak of the flux excess in the PS2 images, for which we have deep reference images
from the 3$\pi$ survey \citep{2016arXiv161205560C}.

\begin{figure*}
    \centering
    \includegraphics[width = 0.95\textwidth]{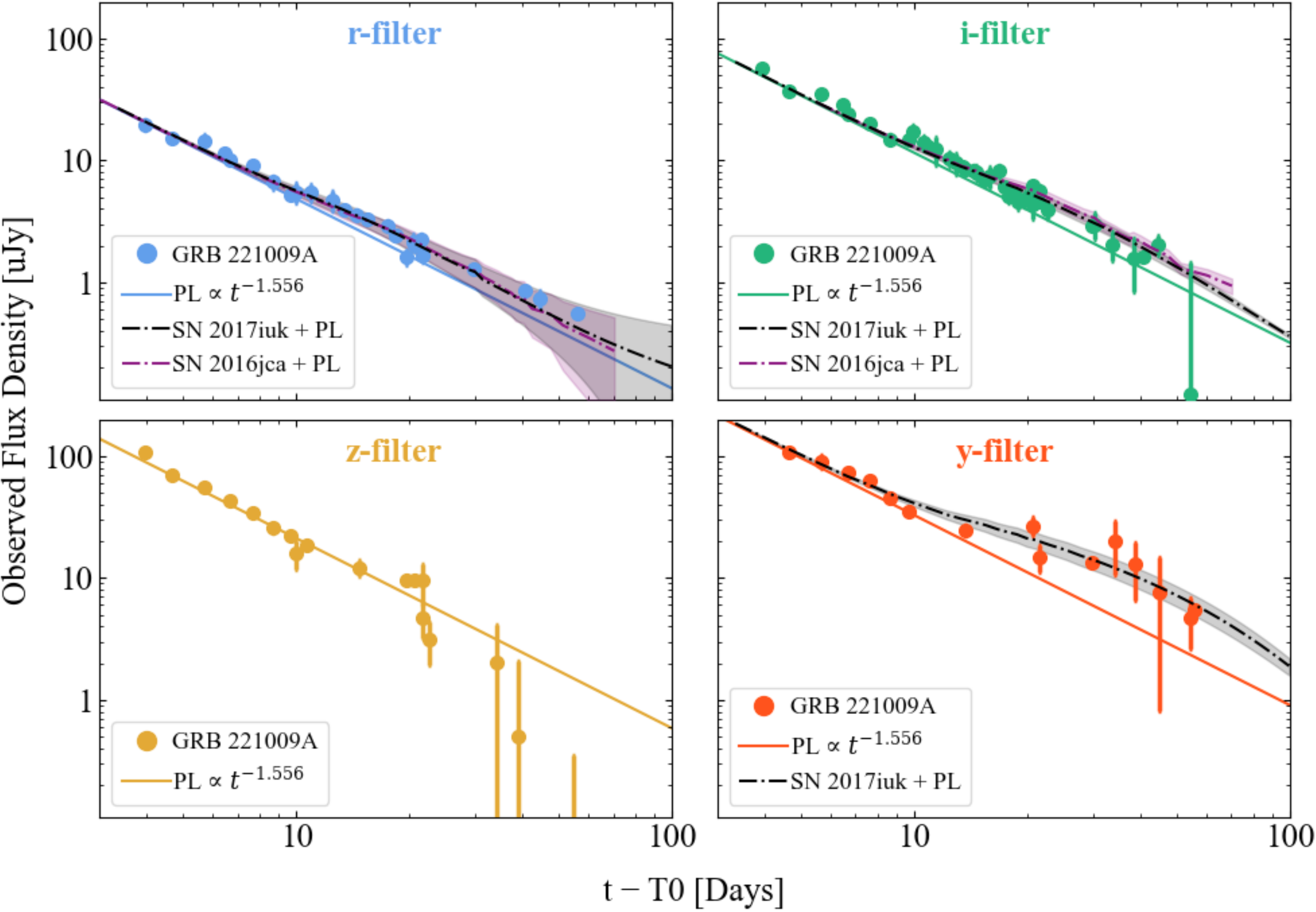}
    \caption{A multi-panel plot of the light curve of GRB~221009A in microJanskys with the X-ray defined power-law fit (PL) of $f(t)\propto t^{-1.556\pm0.002}$. All forced photometry measurements are included, irrespective of significance. The redshifted and reddened light curves of SN~2016jca and SN~2017iuk have been combined with flux from the X-ray power-law model and further stretched in flux to depict the SN signature within GRB~221009A. Shaded regions depict the combined error associated with the SN measurements, power-law modelling, and dust extinction.}
    \label{fig:sn-sig}
\end{figure*}

\section{Interpretation of the excess flux as a supernova signature}
\label{sec:sn}

To determine if the excess flux observed starting from $\sim7-8$ days is consistent with a Type Ic Broad-Lined supernova (Ic-BL) contribution, we compare the excess to the $riz$ light curve of SN~2016jca from \cite{2017A&A...605A.107C} and the $griz$ light curve of SN~2017iuk from \cite{izzo2019GRBSNsignatures}. 
SN~2016jca was a type Ic-BL that emerged in GRB~161219B \citep{2017A&A...605A.107C},  at a redshift of $z=0.1475$. This redshift is almost identical to GRB~221009A, and as such, the $riz$ light curve can be compared directly.
SN~2017iuk was another type Ic-BL that emerged in GRB~171205A. It is the only type Ic-BL with published rest frame $z$-band filter coverage, to compare to the observer frame \yps\ data of GRB~221009A.  At this redshift, the GROND $griz$ SDSS filters \citep{1996AJ....111.1748F} 
 correspond to the observer frame of the Pan-STARRS system $rizy_{\rm PS}$-filters at $z=0.151$ to a good approximation (see Table\,\ref{tab:3Pistacks}). In particular the restframe $z$-band
transforms to the observed \yps\ filter 
 and since the \yps\ data apparently show the strongest excess, the $z-$band restframe data are essential.

We used the observed data of GRB~171205A and SN~2017iuk listed in \cite{izzo2019GRBSNsignatures} which is already corrected for host galaxy contribution, and we further corrected for dust extinction to the SN. We do not subtract off any X-ray-derived power-law as the afterglow contribution represented by which is considerably less than the SN contribution as early as three days after explosion.
As the light curve of 2017iuk is only $\sim$27 days long, we generated a continuous model light curve fit up to 100 days after explosion using an MCMC Bazin fit \citep{2009A&A...499..653B} for each of the $griz$ data. 

To obtain the light curve of SN~2016jca we used the observed data listed in 
\citep{2017A&A...605A.107C} and subtracted off an X-ray-derived power-law of $f(t)\propto t^{-0.79}$ 
fitted to the early-time data (
$t-T_0 <1$\,day) to remove the GRB~161219B afterglow contribution. We also corrected for the host galaxy contribution (using the same host flux determined by \cite{2017A&A...605A.107C}) and dust extinction to the SN.
The light curve of SN~2016jca is well sampled up to ~70 days after explosion, and so we did not require a Bazin fit here.

These light curves were then adjusted accordingly for extinction along the line-of-sight to GRB~221009A and time dilation to the GRB redshift of $z=\Hostz$.  
We used the total estimated line of sight extinction implied from the X-ray to optical power-law slope as derived in Section\,\ref{sec:analysis} (which is higher than that from Milky Way only), explicitly $A_r = \Artot$, $A_i=\Aitot$, $A_z=\Aztot$, $A_y =\Aytot$. 
These light curve fluxes of SN~2016jca and SN~2017iuk were subsequently added to the X-ray power-law slope derived in Section\,3 and overlaid onto the GRB afterglow in Figure\,\ref{fig:sn-sig}. 

The light curve fluxes of SN~2016jca required moderate (arbitrary) scaling to produce a
power-law plus SN component that quite satisfactorily matches the observed data of GRB~221009A. Scaling factors of $2.0$ and $1.6$ were used for the $r$-filter and $i$-filter respectively. We computed Bayes Factors to compare the continuation of the imposed X-ray power-law to the power-law plus an SN component. 
We found factors of $3.4$ and $7.5$ for the $r$-filter and $i$-filter respectively. This favours a model with a power-law plus SN rather than the imposed x-ray power-law only. The light curve shape and peak magnitudes are similar 
(within the errors) to that of SN~2016jca 
\citep{2017A&A...605A.107C,2019MNRAS.487.5824A}. We estimate rest-frame, absolute magnitudes of $M_g=-19.7\pm0.6$ and $M_r=-19.6\pm0.3$ for a Ic-BL SN component of similar nature to SN~2016jca.

The SN~2017iuk light curve model fluxes in the $griz$-filters required considerable scaling (arbitrarily) to match the corresponding observed fluxes of GRB~221009A in the $riz\yps$ bands.
We require scaling factors in the restframe $grz$-filters 
of $4.8$, $1.9$, and $4.6$, respectively, 
to match the excess flux observed in the GRB data. 
This would produce peak magnitudes of 
$M_g=-19.9\pm0.6$ and $M_r=-19.3\pm0.3$ and $M_z=-20.1\pm0.3$
for the SN component inside GRB~221009A. No consistent single
scaling factor can produce the colors and peak magnitudes in all filters. However, the variation in color of type Ic supernovae 
and the uncertain extinction toward GRB~221009A (both Milky Way and the additional
required extinction) may explain why differences are found in each filter. We note that a difference in scaling factors was considered in the \cite{2017A&A...605A.107C} interpretation of SN~2016jca.

We computed Bayes Factors (in the same way as for the SN~2016jca comparison) and found $3.4$, $7.4$ and $3.9$ for the $grz$-filters respectively.
We emphasise that this is comparing the imposed x-ray power-law (no SN) to 
power-law plus SN component after 7 days. Again, this method favours a Ic-BL component, but only with these assumptions.

We do not measure a peak absolute magnitude in the observer frame $z$-filter for either SN comparison, as the Bayes Factors from such comparisons strongly favoured a ``no SN'' solution regardless of the scaling factors used. We also considered a scenario without the extra dust extinction suggested by the X-ray analysis. In this, we measure peak absolute magnitudes of $M_g=-18.7\pm0.6$, $M_r=-18.6\pm0.3$ and $M_z=-19.6\pm0.3$ from the SN~2017iuk comparison and $M_g=-18.6\pm0.6$, $M_r=-18.7\pm0.3$ from the SN~2016jca comparison.

We subtracted the afterglow (X-ray power-law) from the observed data and measured the color of the flux excess at two phases: peak brightness ($t-T_0 \sim 21$\,days) and at late times ($t-T_0 \sim 34$\,days), to check for a color change which may be indicative of an SN component. Using only measurements made through difference imaging to mitigate host contribution, and discounting dust extinction, we measure observed $i-y$ colors ($r-z$ in the rest frame) of $i-y=2.77\pm0.56$ and $i-y=4.24\pm1.95$ at peak brightness and late times, respectively. The relatively large errors imply no meaningful color evolution can be interpreted.

\begin{figure*}
    \centering
    \includegraphics[width = 0.85\textwidth]{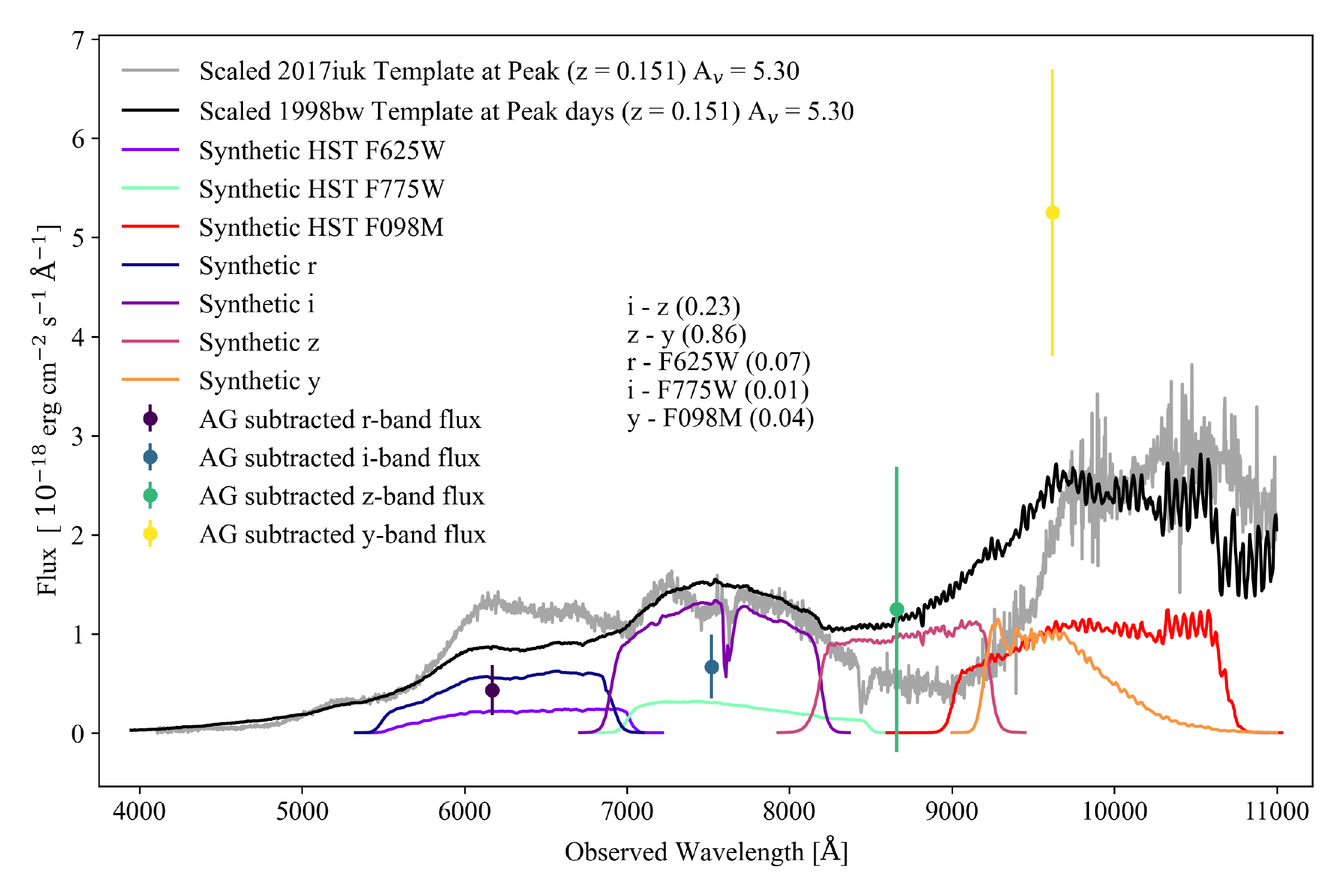}    
    \caption{The spectrum of SN1998bw at peak \citep{2001ApJ...555..900P} reddened with $A_V=5.30$, redshifted to z = 0.151 and scaled to approximate the afterglow subtracted $rizy$ flux at +20 days (after GRB), respectively. The convolution of the SN1998bw template with the transmission profiles of the filters used is overlaid. For comparison, following the same method, a scaled, reddened, and redshifted spectrum of SN 2017iuk at maximum light \citep{izzo2019GRBSNsignatures} is also included.}
    \label{fig:synphot}
\end{figure*}

To investigate why no apparent excess flux is visible in the 
observer frame $z$-filter (rest frame $i$-filter), we took the spectrum of SN1998bw and SN~2017iuk at peak and reddened both with 
the total extinction that we estimated in Section\,\ref{sec:data}
($A_V=5.3$) 
and redshifted it to $z=0.151$. 
These are plotted in Figure\,\ref{fig:synphot} along with the convolution of the spectra with the effective transmission curves of the filters  used  \footnote{Transmission profiles for all filters were obtained from the SVO Filter Profile Service. http://svo2.cab.inta-csic.es/theory/fps}. At $z=0.151$, the strong \CaII\ triplet has the deep 
P-Cygni absorption precisely at the position of the \zps\ filter. The \ips\ filter and the 
\yps\ filter cover the emission peaks of the \CaII\ triplet and the 7500\AA\ blend. This may explain why there is little-to-no sign of a Ic-BL supernova
signature in the $z$-filter but is plausibly detected in the other filters.

We conclude that the excess flux above the extrapolation of the afterglow power-law of $f(t)\propto t^{-1.556\pm0.002}$ can be explained by the 
emergence of a supernova similar in luminosity and duration to the observed type Ic-BL SNe SN~2016jca and SN~2017iuk.  
This supernova associated with GRB~221009A would be labelled  SN~2022xiw as reported on
the IAU Transient Name server \citep{2022TNSCR3047....1P}. 
At the later epochs, our PS2 images have a depth and sensitivity close to those of the reference 
images we use for image subtraction. We demonstrate the reality of the detections compared with the PS1 3$\pi$ stacks  at several important epochs in Figure\,\ref{fig:images}.

We compare our $r$ and $i$-filter photometry with original data (not GCN values) in \cite{2023arXiv230204388L}, \cite{2023arXiv230203829S} and \cite{2023arXiv230207761L}, three papers which suggest there is no strong evidence for a supernova
component in the data. Their data are in excellent agreement with ours, even though no image
subtraction was undertaken.  
The $z$-filter data of \cite{2023arXiv230204388L} and \cite{2023arXiv230203829S} are slightly 
brighter than ours after 8-10 days, which may suggest some host galaxy contribution affects their measurements. The converted $y$-filter from the HST data of \cite{2023arXiv230207761L} are fainter than ours by $\sim0.3$ magnitudes at the respective epochs, which we attribute to the combined galaxy plus PSF fits of \cite{2023arXiv230207761L}.
Neither group have extensive $y$-filter or photometric NIR data at $>$\,10 days. While 
\cite{2023arXiv230206225K} also claim no evidence for (or against) a supernova component, 
their measurements include only 2 epochs of their own data beyond 10 days and no meaningful comparison is possible with our data set.

It appears that an SN component is only detectable with extensive and accurate photometric coverage beyond 10 days and assuming the X-ray power-law is applied to the optical data \citep[which may be disputed e.g.][]{2023arXiv230204388L}. 
If we were to assume there is no SN component, and that there is a break in the afterglow SED between the X-ray and optical decline rates, then, by applying similar extinction corrections, we find almost identical agreement with \cite{2023arXiv230204388L}, \cite{2023arXiv230203829S} and \cite{2023arXiv230207761L} in the power-law decline rates for the $riz$-filters across all epochs (see top panel of Figure\,\ref{fig:afterglow-lc}), even though our data is better sampled and extends further in time. However, it is not possible to fit our $y$-filter with the same single decaying power-law derived from the $riz$-filters, which would imply the achromatic behaviour in the afterglow. Although, we acknowledge that if we were to replace our HST $y$-filter measurements with that reported in \cite{2023arXiv230207761L}, then we find it possible to describe the $y$-filter decay with the same power-law derived from the $riz$-filters. We measure this single decaying power-law as $f(t)\propto t^{-1.46 \pm 0.05}$.

\section{Discussion and Conclusions}
\label{sec:discuss} 

Given that the observational data of GRB~221009A and the emergence of a 
supernova (SN~2022xiw) could be explained by the addition of scaled observed fluxes
of two type Ic-BL SNe, we can determine the physical parameters of SN~2022xiw. We model the afterglow subtracted SN signal to constrain 
the ejecta mass ($M_{\rm ej}$), nickel mass ($M_{\rm Ni}$) and ejecta expansion velocity 
($v_{\rm ej}$) using \texttt{MOSFiT}, an open-source, one-zone, semi-analytical fitting tool for 
broadband SN light curves \citep{guillochon2018mosfit}. \texttt{MOSFiT} employs \texttt{dynesty}, a nester sampling technique \citep{speagle2020dynesty}, to estimate posteriors of the fitted model. We use the default  
Arnett model within \texttt{MOSFiT}, which assumes that the SN is entirely powered by the radioactive decay of $^{56}$Ni and $^{56}$Co and 
assumes that the spectral energy distribution can be described by a black body \citep{arnett1982type,villar2017theoretical}.
Assuming a constant grey opacity 
$\kappa=0.07$\,cm$^{2}$g$^{-1}$ 
\citep{2000AstL...26..797C,2017A&A...605A.107C}
and flat priors on all fitted parameters, we find the posterior distributions returned are 
$M_{\rm ej}=5.6^{+1.1}_{-0.9}M_{\odot}$, 
$M_{\rm Ni}=0.5^{+0.20}_{-0.1}M_{\odot}$ and 
$v_{\rm ej}=35,500^{+4,300}_{-4,600}$\,\kms\ 
to model the $riz$-band data of SN~2016jca (see Appendix\,\ref{app:mosfitfigs} Figure\,\ref{fig:mosfitfigs-16jca}). Errors quoted represent a $1\sigma$ width of the posterior distributions.
The inferred explosion energy is then $E_{\rm kin}\simeq2.7-6.3\times10^{52}$\,ergs. 
These are broadly similar to those derived 
in the previous analysis of \cite{2017A&A...605A.107C} and \cite{2019MNRAS.487.5824A}. Our $M_{\rm Ni}$ is higher, which we attribute to the different methods 
of fitting a bolometric light curve rather than the black body 
fitting of the selected filters. 

Using the same set of priors, we model the $riy$-filter light curve of SN~2022xiw. The posterior distributions return the values 
$M_{\rm ej}=7.1^{+2.4}_{-1.7}$\,M$_{\odot}$, 
$M_{\rm Ni}=1.0^{+0.6}_{-0.4}$\,M$_{\odot}$,
and 
$v_{\rm ej}=33,900^{+5,900}_{-5,700}$\,\kms, 
(see Figure\,Appendix\,\ref{app:mosfitfigs}). 
inferring explosion energy of 
$E_{\rm kin}\simeq 2.6-9.0\times10^{52}$\,ergs. 
These are comparable to the quantities derived for SN~2016jca above
but suggest a more energetic explosion and a larger mass of
$^{56}$Ni is required. This is reflected in the fact that the light curve data of SN~2016jca requires modest scaling to 
reproduce the flux of SN~2022xiw.  We test the robustness of our modelling by excluding the \yps-band photometry (which may significantly deviate from our black body assumptions) and by excluding late-time observations (which may deviate from our assumption that the SN is in the photospheric phase). We find no statistically significant differences in derived properties from these tests. 

The parameters of SN~2022xiw 
are similar to the more energetic Ic-BL SNe associated with lGRBs,  
termed hypernovae \citep{1998Natur.395..672I,2003ApJ...599L..95M}. The sample of GRB-SNe has been reviewed 
and summarised by \cite{2012grb..book..169H} and more recently \citep{2015MNRAS.450.1295W} and \cite{2017AdAst2017E...5C}.
The latter studies suggest that GRB-SNe are characterised by the following average values: 
kinetic energies of 
$\overline{E_{\rm K}}=2.5\times10^{52}$\,erg ($\sigma_{\rm E_K}=1.8\times10^{52})$, 
ejecta masses of 
$\overline{M_{\rm ej}}=6$\msol\, ($\sigma_{\rm M_{\odot}}=4$\,\msol). 
and peak photospheric velocities of 
$\overline{v_{\rm ph}}=20,000$\msol\, ($\sigma_{\rm v_{ph}}=8000$\,\msol). The most luminous 
GRB-SNe require $^{56}$Ni masses of $0.5 \lesssim M_{\rm Ni} \lesssim 0.9$, if the 
luminosity is powered by radioactivity. 

The $^{56}$Ni mass we derive ($M_{\rm Ni}=1.0^{+0.6}_{-0.4}$\,M$_{\odot}$), is on the high side of the known distribution of GRB-SNe. Similar to the SN~2016jca case above, we attribute this partly to fitting a computed bolometric light curve, which is made worse by the bright afterglow of GRB~221009A and the high, uncertain extinction creating considerable noise in the SN extracted flux. Nevertheless, the large uncertainty comfortably brackets GRB-SNe parameters previously derived. 

Despite these uncertainties, the excess flux above the extrapolated afterglow leads us to conclude that there may be  a supernova signature in our $\sim60$\,day, well-sampled light curve data of GRB~221009A (denoted SN~2022xiw) which is comparable in both luminosity and ejecta properties to other bright type Ic-BL supernovae events. To confirm, this will require a reanalysis of all multi-wavelength data, to model the expected optical afterglow behaviour.

\section{acknowledgments}
Pan-STARRS is a project of the Institute for Astronomy of the University of Hawaii, and is supported by the NASA SSO Near Earth Observation Program under grants 80NSSC18K0971, NNX14AM74G, NNX12AR65G, NNX13AQ47G, NNX08AR22G, 80NSSC21K1572 and by the State of Hawaii.
The Pan-STARRS1 Sky Survey data were facilitated by  the University of Hawaii, the Pan-STARRS Project Office, the Max Planck Society 
(MPIA, MPE), Johns Hopkins University, Durham University,  University of Edinburgh,  Queen's University Belfast, the Harvard-Smithsonian CfA the Las Cumbres Observatory Global Telescope Network Incorporated, the National Central University of Taiwan, the Space Telescope Science Institute, NASA Grant No. NNX08AR22G, 
NSF Grant No. AST-1238877, the University of Maryland, Eotvos Lorand University, and the Los Alamos National Laboratory.  
The Young Supernova Experiment (YSE) and its research infrastructure is supported by the European Research Council under the European Union's Horizon 2020 research and innovation program (ERC Grant Agreement 101002652, PI K.\ Mandel), the Heising-Simons Foundation (2018-0913, PI R.\ Foley; 2018-0911, PI R.\ Margutti), NASA (NNG17PX03C, PI R.\ Foley), NSF (AST-1720756, AST-1815935, PI R.\ Foley; AST-1909796, AST-1944985, PI R.\ Margutti), the David \& Lucille Packard Foundation (PI R.\ Foley), VILLUM FONDEN (project 16599, PI J.\ Hjorth), and the Center for AstroPhysical Surveys (CAPS) at the National Center for Supercomputing Applications (NCSA) and the University of Illinois Urbana-Champaign.
S. J. Smartt, K. W. Smith and D. Young acknowledge STFC grants ST/P000312/1. 
L. Izzo and J. Hjorth were supported by a VILLUM FONDEN Investigator grant awarded to J. Hjorth (project number 16599).
C. D. Kilpatrick was supported in part by a CIERA Postdoctoral Fellowship.
M. Nicholl is supported by the European Research Council (ERC) under the European Union’s Horizon 2020 research and innovation programme (grant agreement No.~948381) and by a Fellowship from the Alan Turing Institute.
S. Yang has been supported by the research project grant "Understanding the Dynamic Universe" funded by the Knut and Alice Wallenberg Foundation under Dnr KAW 2018.0067, and the G.R.E.A.T research environment, funded by {\em Vetenskapsr\aa det}, the Swedish Research Council, project number 2016-06012.
D. A. Coulter acknowledges support from the National Science Foundation Graduate Research Fellowship under Grant DGE1339067.
The UCSC team is supported in part by NASA grant 80NSSC20K0953, NSF grant AST--1815935, the Gordon \& Betty Moore Foundation, the Heising-Simons Foundation, and by a fellowship from the David and Lucile Packard Foundation to R. J. Foley.
This work includes data obtained with the Swope Telescope at Las Campanas Observatory, Chile, as part of the Swope Time Domain Key Project (PI: Piro, Co-Is: Drout, Phillips, Holoien, French, Cowperthwaite, Burns, Madore, Foley, Kilpatrick, Rojas-Bravo, Dimitriadis, Hsiao). We thank Abdo Campillay and Yilin Kong Riveros for performing the Swope observations.
Data from NOIR DECam supported by proposals 2021A-0275 and 2021B-0325.
We thank Lulin staff H.-Y. Hsiao, C.-S. Lin, W.-J. Hou, H.-C. Lin and J.-K. Guo for observations and data management. T.-W. Chen thanks D. B. Malesani for photometric measurement on the SLT GCN circular and Y.-C. Cheng for LOT time. 
We acknowledge use of NASA/IPAC Extragalactic Database (NED),  operated by JPL, Caltech under contract with NASA. 
%
%

\vspace{5mm}
\facilities{Pan-STARRS, Swift(XRT), CTIO:4m,DECam,
CFHT(MEGACAM), SOAR, HST(WFC3), Swope-1m, Liverpool Telescope (IO:O), Lulin:0.4m-SLT, 1m-LOT.}


\software{
Astropy \citep{2013A&A...558A..33A,2018AJ....156..123A,2022ApJ...935..167A},
DOLPHOT \citep{2016ascl.soft08013D},
MOSFiT \citep{guillochon2018mosfit},
Photpipe \citep{2005ApJ...634.1103R,2007ApJ...666..674M,Rest_2014}
Photutils \citep{larry_bradley_2022_6825092},
Scipy \citep{2020SciPy-NMeth},
YSE-PZ \citep{2022zndo...7278430C}
}



\appendix

\section{Photometry data table}
\label{app:photometrytable}

A CSV file is provided, with all the photometry data for GRB~221009A behind Figure\,\ref{fig:afterglow-lc} and Figure\,\ref{fig:sn-sig}. Measurements are in AB magnitudes and microJanskys. All photometry is of the AG+SN and are uncorrected for galactic and host galaxy dust extinction. Magnitude limits are quoted to 2-sigma. The total exposure time for each night is provided which is typically the sum of sub-exposures combined. 
 The final column notes the method used to measure the fluxes for that particular epoch i.e. stacked fluxes from individual difference images (\textit{w-avr-flx}), stacked fluxes from individual difference images binned across multiple nights (\textit{bw-avr-flx}), fluxes from a single stacked difference image (\textit{d-stack-flx}) or fluxes from a single stacked target image without differencing (\textit{target-flx)}.

 \section{Posterior distributions for the physical parameters of SN~2016jca and SN~2022xiw}
 \label{app:mosfitfigs}
 
 \begin{figure}
    \centering
    \includegraphics[width = 0.60\textwidth]{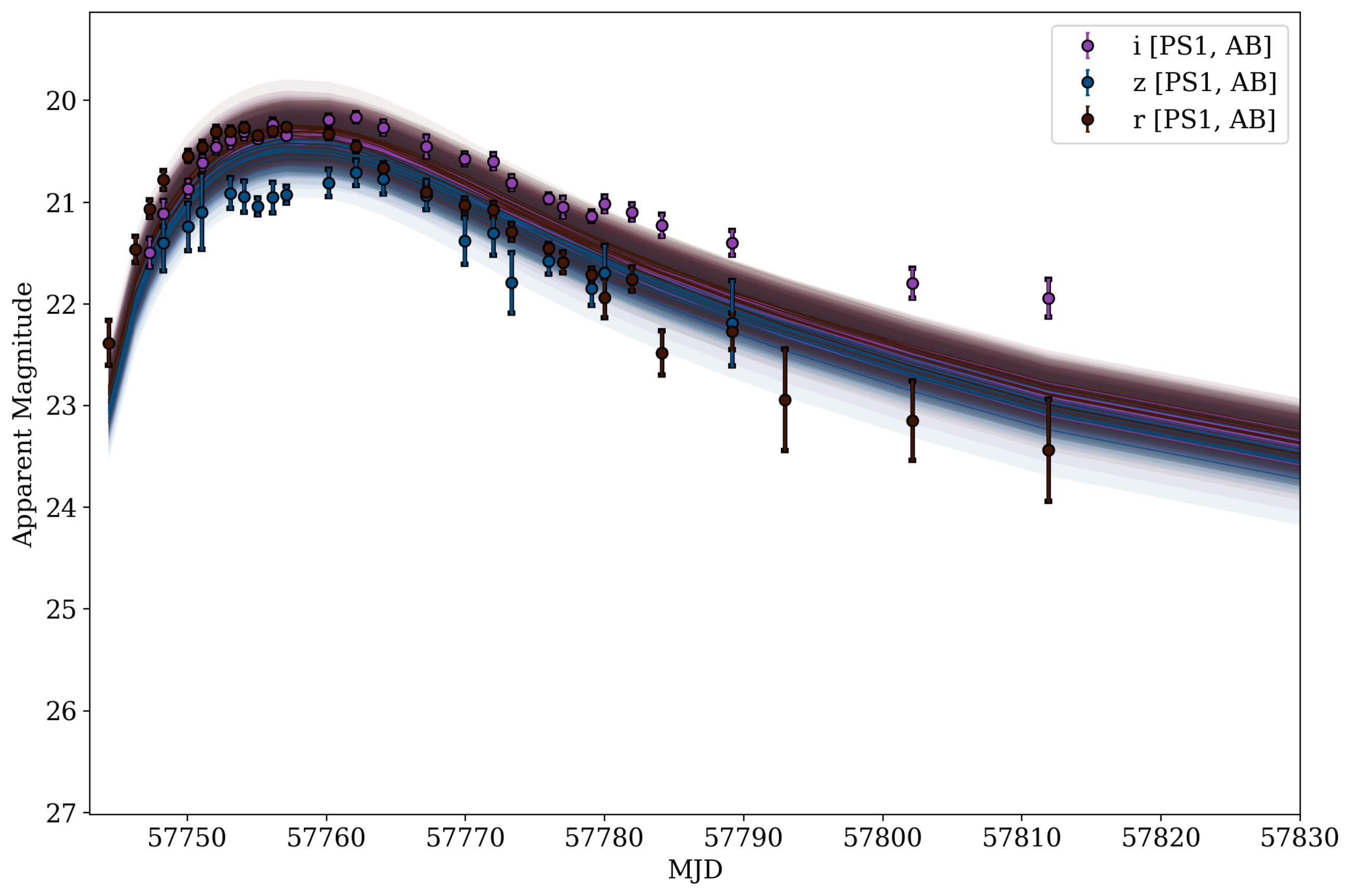}
    \includegraphics[width = 0.75\textwidth]{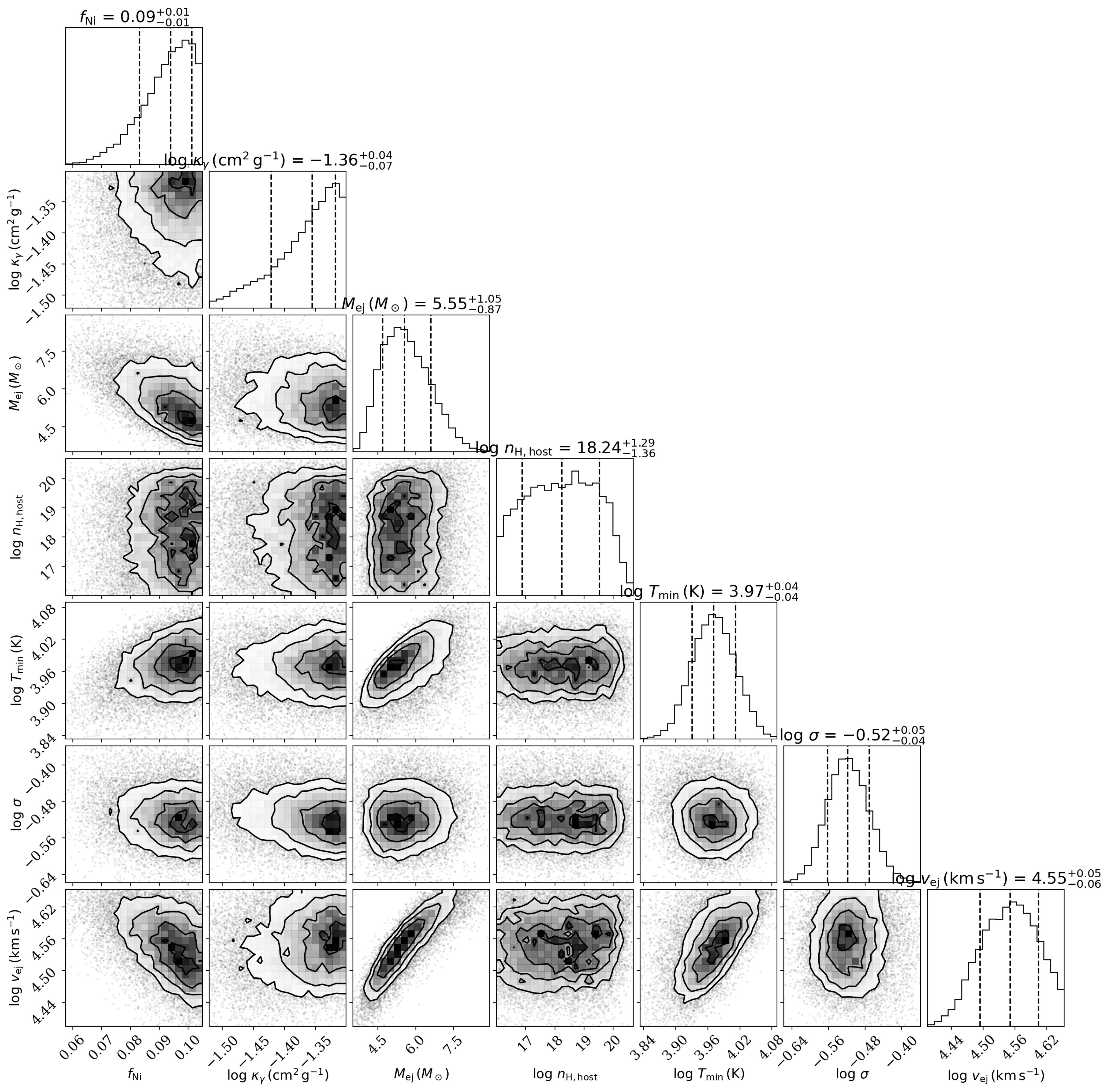}
    \caption{\textit{Upper Panel}: MOSFiT light curve models for SN~2016jca.
    \textit{Lower panel}: Derived parameters from the MOSFiT modelling.}
    \label{fig:mosfitfigs-16jca}
\end{figure}

 \begin{figure}
    \centering
    \includegraphics[width = 0.60\textwidth]{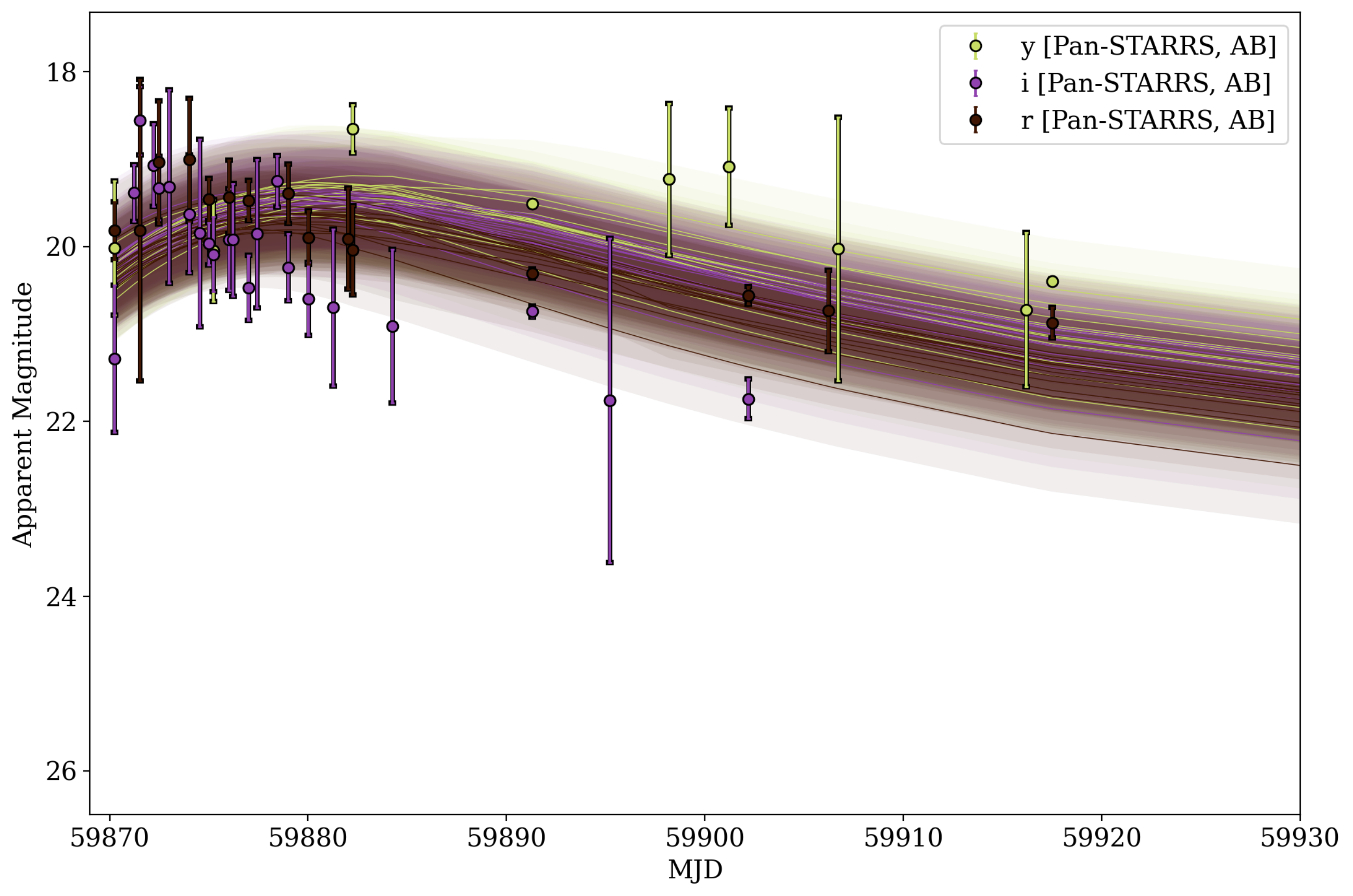}
    \includegraphics[width = 0.75\textwidth]{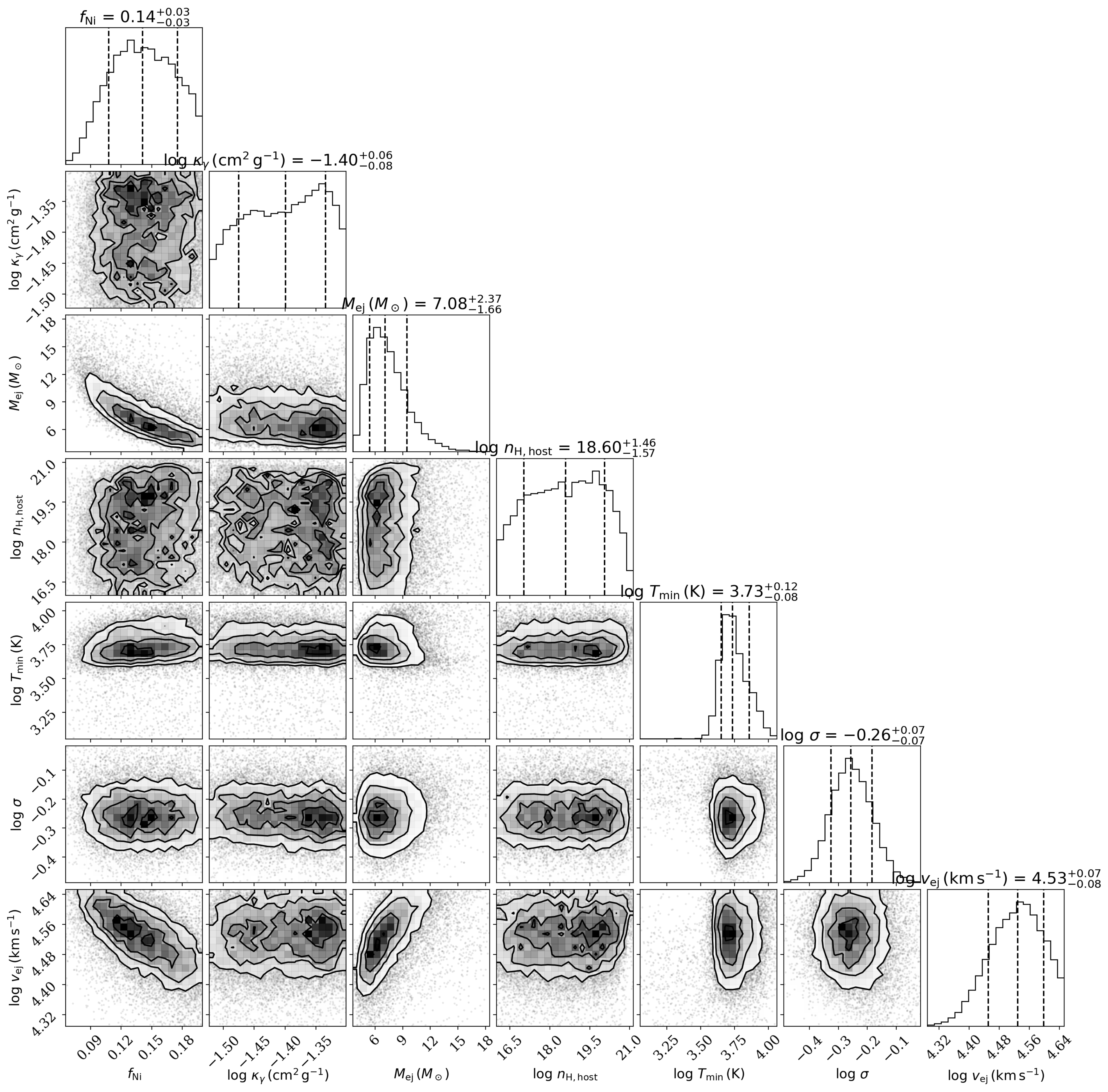}
    \caption{\textit{Upper Panel}: MOSFiT light curve models for SN~2022xiw.
    \textit{Lower panel}: Derived parameters from the MOSFiT modelling.}
    \label{fig:mosfitfigs-22xiw}
\end{figure}


\bibliography{main}{}
\bibliographystyle{aasjournal}



\end{document}